\documentclass[
 reprint,
superscriptaddress,
 amsmath,amssymb,
 aps,
floatfix,
]{revtex4-2}

\usepackage{graphicx}% Include figure files
\usepackage{dcolumn}% Align table columns on decimal point
\usepackage{bm}% bold math
\usepackage{xcolor}
\begin{document}

\preprint{APS/123-QED}

\title{XUV fluorescence as a probe of laser-induced helium nanoplasma dynamics}
%\thanks{A footnote to the article title}

\author{Malte Sumfleth}
\affiliation{Deutsches Elektronen-Synchrotron DESY, Notkestr. 85, 22607 Hamburg, Germany}
\author{Andreas Przystawik}
\affiliation{Deutsches Elektronen-Synchrotron DESY, Notkestr. 85, 22607 Hamburg, Germany}
\author{Mahesh Namboodiri}
\affiliation{Deutsches Elektronen-Synchrotron DESY, Notkestr. 85, 22607 Hamburg, Germany}
\author{Tim Laarmann}
\email{tim.laarmann@desy.de}
\affiliation{Deutsches Elektronen-Synchrotron DESY, Notkestr. 85, 22607 Hamburg, Germany}
\affiliation{The Hamburg Centre for Ultrafast Imaging CUI, Luruper Chaussee 149, 22761 Hamburg, Germany}

%\collaboration{MUSO Collaboration}%\noaffiliation

%\author{Charlie Author}
% \homepage{http://www.Second.institution.edu/~Charlie.Author}
%\affiliation{
% Second institution and/or address\\
% This line break forced% with \\
%}%
%\affiliation{
% Third institution, the second for Charlie Author
%}%
%\author{Delta Author}
%\affiliation{%
% Authors' institution and/or address\\
% This line break forced with \textbackslash\textbackslash
%}%

%\collaboration{CLEO Collaboration}%\noaffiliation

\date{\today}% It is always \today, today,
             %  but any date may be explicitly specified

\begin{abstract}
XUV fluorescence spectroscopy provides information on energy absorption and dissipation processes taking place in the interaction of helium clusters with intense femtosecond laser pulses. The present experimental results complement the physical picture derived from previous electron and ion spectroscopic studies of the generated helium nanoplasma. Here, the broadband XUV fluorescence emission from high-lying Rydberg states that covers the spectral region from $6p \to 1s$ at 53.0\,eV all the way to photon energies corresponding to the ionization potential of He$^+$ ions at 54.4\,eV is observed directly. The cluster size-dependent population of these states in the expanding nanoplasma follows the well-known bottleneck model. The results support previous findings and highlight the important role of Rydberg states in the energetics and dynamics of laser-generated nanoplasma.      
\end{abstract}

%\keywords{Suggested keywords}%Use showkeys class option if keyword
                              %display desired
\maketitle

%\tableofcontents

\section{\label{sec:Introduction} Introduction}
The interaction of strong laser fields with nanoparticles generating plasma at the nanoscale have enjoyed tremendous attention from researchers throughout the last couple of decades\,\cite{KRAINOV2002237,Saalmann_2006,RevModPhys.82.1793,C3CP55380A}. Numerous studies have investigated energy deposition and redistribution in atomic and molecular clusters as a function of cluster size situating them between atoms and bulk matter with dimensions down to a few nanometers. Using cluster beams as target material combines the advantage of solid state density of the individual particles with the low target density in the interaction volume. This experimental scheme creates well-defined conditions for studying nanoplasma formation and its relaxation across a large range of excitation energies from the IR with laser wavelengths $\lambda >1$\,$\mu$m\,\cite{Schuette2016}, across the VUV\,\cite{Nature.420.482,SCHULZ2003572,PhysRevLett.92.143401,PhysRevLett.95.063402,Ziaja_2009,PhysRevA.100.063402} and XUV spectral range\,\cite{Hoener_2008,PhysRevLett.108.093401} towards x-rays at $\lambda <10$\,nm\,\cite{PhysRevLett.108.133401, PhysRevLett.108.245005}. 

Massive electron excitation and ionization generates a deep mean-field potential on the order of keV energy confining the plasma electrons. Enormous electron temperatures beyond $10^6$\,K can be reached by different mechanisms of energy absorption by the quasifree plasma electrons, which mainly depend on the laser parameters and the particle size. Efficient energy deposition results in the ejection of hot electrons \cite{PhysRevA.68.053201}, highly charged\,\cite{PhysRevLett.77.3347} and energetic ions\,\cite{PhysRevLett.80.261}, as well as x-ray emission from the nanoplasma\,\cite{McPherson1994}. Even nuclear fusion in laser-heated deuterium clusters has been observed\,\cite{Ditmire1999}. 

The optimization of secondary electron, ion, x-ray or pulsed neutron sources towards applications requires detailed knowledge on the evolution of the electronic properties of the expanding nanoplasma driven by the Coulomb forces of ions and the hydrodynamic pressure of electrons. Experimental studies flanked by theoretical calculations show that the electron thermalization to the bottom of the plasma potential is fast. It takes place on a fs timescale due to the bulk-density conditions and is accompanied by isotropic emission of low-energy (thermal) electrons\,\cite{PhysRevLett.121.063202}. During the nanoplasma expansion the binding energy of the quasifree electrons gradually shifts upwords forming a delocalized energy band close to the ionization continuum. In this phase, electron-correlation-driven energy transfer processes have to be considered\,\cite{Schuette2015,Oelze2017,Niozu2019}. In parallel, the mean-field potential upshift and the resulting increase in the interatomic Coulomb barriers causes localized states of individual ions to recover\,\cite{PhysRevLett.112.253401}. Thereby, triggered by electron-ion recombination processes various additional electronic excitation and energy relaxation channels open up on the picosecond (ps) to nanosecond (ns) time scale\,\cite{NewJPhys.15.053047,NewJPhys.17.033043} including radiative decay\,\cite{PhysRevLett.112.183401,Mueller2015,Przystawik2015}. This is where the present fluorescence spectroscopy study comes into play to shed light on the less-explored energy redistribution processes in the long-time evolution of a laser-generated helium nanoplasma \,\cite{doi:10.1063/1.5089943,PhysRevLett.125.093202}.

Different types of recombination pathways are discussed in the literature\,\cite{Yukap_Hahn_1997}. In the radiative recombination of an electron with an ion its excess energy is released as a photon. Three-body recombination (TBR) processes involve two electrons and one ion. One of the two electrons recombines with the ion, while the other electron takes away the excess energy of the first electron. Dielectric recombination describes a variant, where the excess energy is taken up by a second electron already present in the ion, which gets lifted into an electronically excited state. Subsequently, the excited electron relaxes back into its ground state via radiative- or Auger-decay. We note in passing that there exist more complicated recombination processes, such as radiative dielectric recombination, where the radiative decay of the excited electrons happens simultaneously with the initial recombination. In the present work, the focus is on radiative recombination and three-body recombination, because the target is the recombination of quasifree electrons with He$^{2+}$ in a fully ionized He nanoplasma forming He$^{+}$. Therefore, no other electrons exist in the ion that could participate in dielectric recombination or more complicated processes.

Investigations of the three-body recombination process started back in the 1960s\,\cite{PhysRev.181.275} and resulted in the formulation of the so-called classical bottleneck model. The defining features of this model are: (a) the temperature dependence for the total recombination rate, which scales as $\propto T^{-9/2}$, and (b) the electronic state $n'$ most likely being populated is the one with $E_{n'}\sim k_{B}T$, because its ionization energy is comparable to the kinetic energy of the thermal quasifree electrons. However, for a cold plasma $T<1000$\,K the bottleneck model no longer applies. It neglects quantum effects, such as the increased de Broglie wavelengths of thermal electrons at the bottom of the plasma potential, which enhances the recombination probability and allows for population of states with ionization energies significantly higher than the bottleneck prediction. Furthermore, high $l$-states of the same $n$ are populated much more frequently in low-temperature plasma compared to the bottleneck regime, where mainly low $l$-states get populated as outlined in \cite{PhysRevLett.98.133201}. In the present contribution we apply orbital-sensitive XUV fluorescence spectroscopy relying on dipole selection rules to gain information on laser-generated He nanoplasma dynamics. Our results underline the importance of three-body recombination processes in the energy dissipation during the plasma expansion in agreement with previous works based on electron and ion spectroscopy\,\cite{doi:10.1063/1.5089943,PhysRevLett.125.093202}. Important for these seminal papers on the temporal development of a laser-induced helium nanoplasma, as well as for the present study, is the fact that helium atoms represent one of the simplest atomic systems. In particular, in hydrogen-like He$^{+}$, the electronic states are energetically degenerate by the orbital momentum quantum number $l$ and electronic transitions result in only a very few fluorescence lines at well-distinguishable photon energies. Fluorescence spectroscopy has been employed before to study laser-induced nanoplasma in heavy rare gas clusters and cluster mixtures comprising more than one element\,\cite{PhysRevLett.112.183401,Mueller2015,Przystawik2015}. Due to the high number of electrons for heavy rare gases clusters, such as Xe$_{N}$, typically a broad charge state distribution is produced in strong laser fields. However, with respect to correlated many-body electronic decay processes (CED), such as three-body recombination, the characteristic fluorescence from electronically excited states overlap and cannot be experimentally resolved. Hence, the advantage of studying systems like Xe$_{N}$ to monitor dependencies of extreme charging on the laser pulse properties becomes a disadvantage when studying CED, making experiments on the lighter element helium extremely appealing. Here, the complementary fluorescence data provide evidence that the population of electronically excited states in the transient He nanoplasma follows the bottleneck model.  

\section{\label{sec:Experimental Setup} Experimental Setup}
\begin{figure}[ht]
\includegraphics[width = 8.7cm]{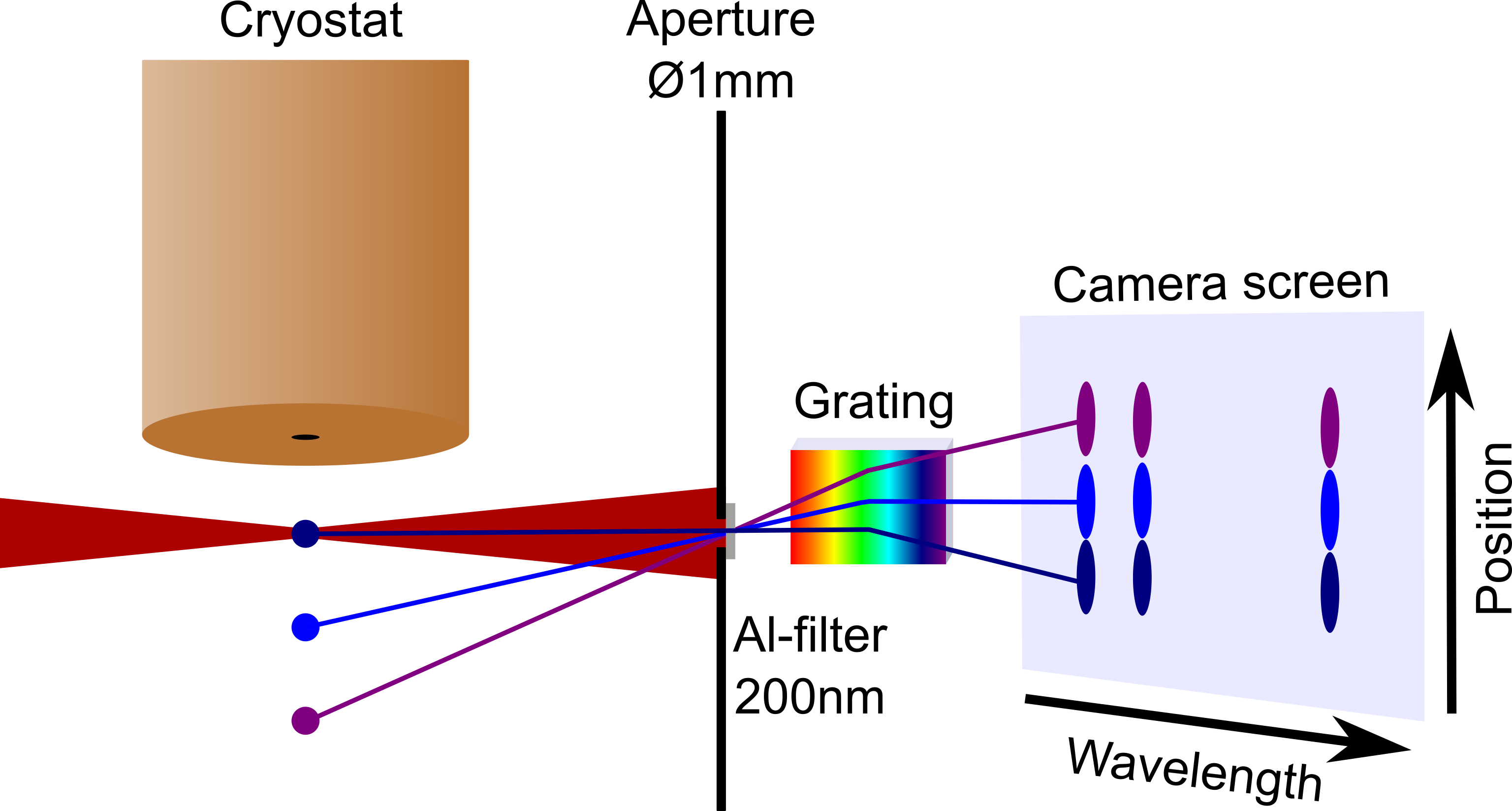}
\caption{\label{fig:camera-screen} The near-infrared laser beam (red) is focused onto the helium droplets generating a nanoplasma. The emitted XUV fluorescence light is imaged vertically through a pinhole and dispersed horizontally by a grating on a XUV camera chip. Fluorescence emitted from different positions along the droplet beam axis is projected to different locations on the detector and provides information on the population dynamics of electronically excited states, their radiative lifetime  and the kinetics of the nanoplasma expansion. The combination of an aluminium filter and the pinhole ensures that the optical laser and stray light is efficiently blocked.}
\end{figure}

An overview of the experimental geometry including the key components is given in Fig.\,\ref{fig:camera-screen}. We generate helium clusters by expanding precooled He gas with a purity of 99.9999\% from 20\,bar stagnation pressure through a circular nozzle into vacuum at temperatures between 4\,K and 15\,K. The nozzle is formed by an electron microscope aperture plate with a 5\,µm orifice, while cooling is provided by a Gifford-McMahon closed-cycle cryostat. This type of He cluster source is well characterized in terms of the temperature and pressure dependent cluster size distribution\,\cite{doi:10.1063/1.3650235}. By changing the nozzle temperature the average particle size can be tuned from rather small He clusters $\overline{N}=3.3\cdot10^3$ with a diameter on the order of a few nm up to large $\mu$m-size droplets comprising more than $10^{10}$ He atoms. The precision of the temperature control is $\pm 30$\,mK, which translates into a size variation, which is much smaller than the full-width at half-maximum of the generated cluster size distribution in a supersonic expansion. The latter is on the order of the average cluster size $\overline{N}$. A few millimeter downstream of the nozzle intense fs laser pulses overlap perpendicularly with the molecular beam.  

A commercial laser system provides the near-IR pulses with a central wavelength of 1030\,nm and a pulse energy of 1\,mJ at a repetition rate of 5\,kHz. The pulse duration of $\tau=180$\,fs has been characterized by measuring the intensity autocorrelation. The fs pulses are focused onto the He clusters by an $f$\,=\,30\,mm lens resulting in a peak intensity beyond 10$^{16}$\,W/cm$^2$. Note, the intensity threshold for barrier suppression ionization (BSI) of individual He atoms is $\approx 10^{15}$\,W/cm$^2$, while BSI in clusters requires a slightly lower laser intensity\,\cite{PhysRevLett.63.2212,PhysRevLett.105.053401}. Thus, rather small clusters with an average size $\overline{N}<10^4$ are almost completely inner-ionized, i.e. the electrons are ionized to a large extent from the individual cluster atoms but do not leave the cluster as a whole, because of the deep plasma potential building up.

Any radiation emitted from the illuminated cluster target in the focal volume has to pass an 1\,mm diameter pinhole at a distance of 284\,mm from the focus and through a 200\,nm-thick aluminum foil, which prevents infrared or visible light from entering the spectrometer chamber in the direction of the laser beam propagation further downstream. In addition uncorrelated stray light scattered from metal surfaces in the source chamber is efficiently blocked at the pinhole. The remaining XUV radiation is dispersed by a toroidal variable line-space grating in the spectrometer chamber onto an XUV camera. 

At first glance, the spectrometer configuration allows us to detect XUV fluorescence photons emitted in the direction perpendicular to the He cluster jet and parallel to the laser beam propagation. However, it is important to note that the setup with the 1\,mm aperture forms a simple pinhole camera as displayed in Fig.\,\ref{fig:camera-screen}. This capability allows to obtain some spatial resolution in the direction of the cluster jet propagation. The location of the detected radiation on the XUV camera chip perpendicular to the dispersion direction of the grating is directly linked to the point in space, where the radiation was emitted. This source point of fluorescence from the expanding He nanoplasma depends on different characteristic velocities and involved time scales, respectively. Clearly, the initial cluster jet speed is small compared to Coulomb explosion and hydrodynamic expansion velocities of the nanoplasma and can be neglected\,\cite{doi:10.1063/1.3650235}. Thus, the location of emitting ions imaged by the pinhole camera mainly depends on the cluster disintegration, the population dynamics of electronically excited states via three-body recombination and their radiative lifetime.

\section{\label{sec:Results} Results and Discussion}
\begin{figure}[ht]
\includegraphics[width = 8.7 cm]{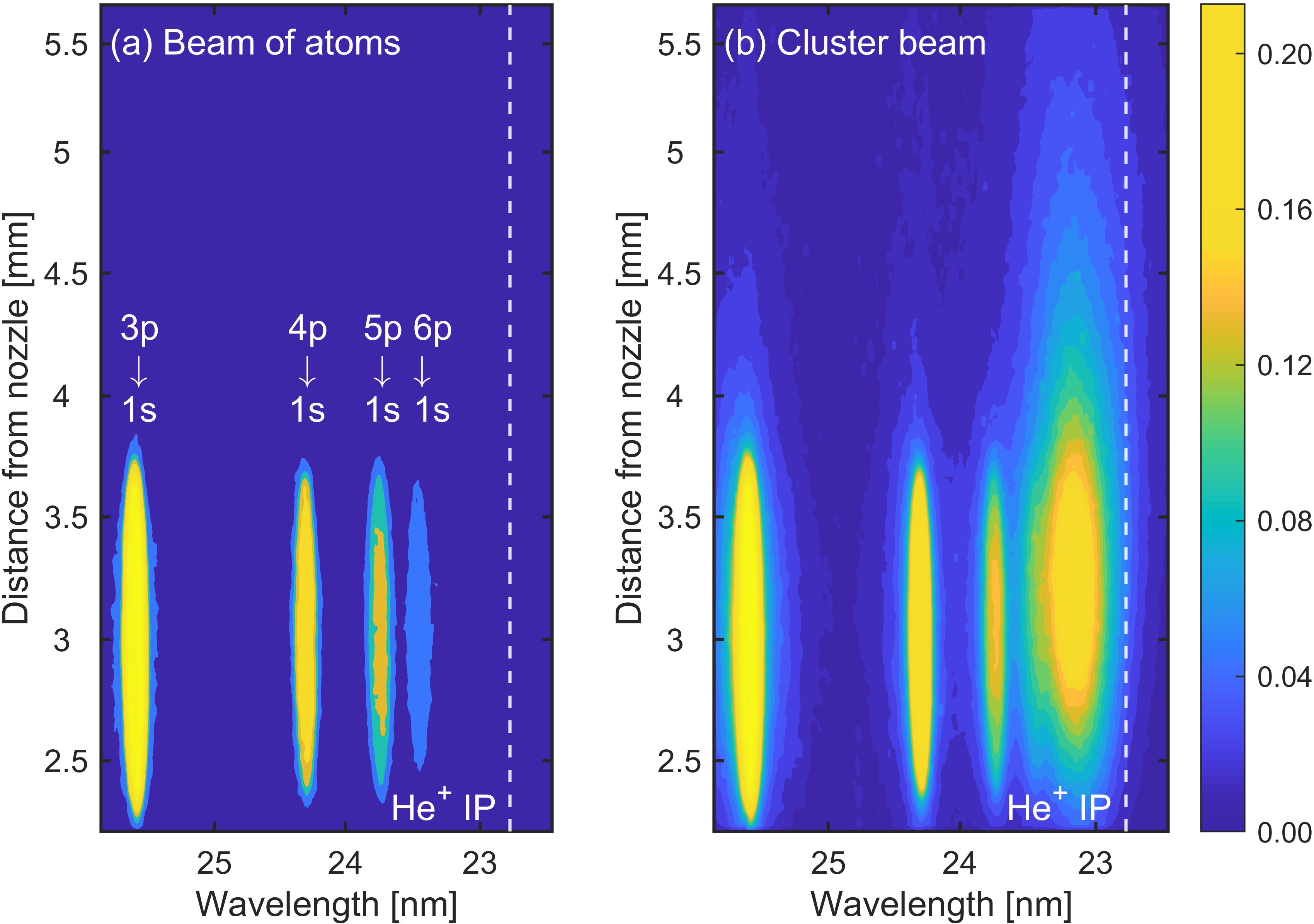}
\caption{\label{fig:Z-Scan-Distribution}Comparison of the recorded XUV-fluorescence spectra using the pinhole camera, when illuminating (a) atomic helium and (b) clusters comprising on average \mbox{$1.2\times 10^6$\,atoms}, which is equivalent to an average particle diameter of about 40\,nm. The positions of vertical arrows mark the atomic transition energies\,\cite{doi:10.1063/1.3077727}. The spectra are normalized to the maximum intensity. The given colour scale applies to both spectra.}
\end{figure}

As the starting point of the present study, we compare the radiative decay of ionized atoms and clusters in Fig.\,\ref{fig:Z-Scan-Distribution}. The experimental procedure is straightforward. It utilizes the fact that the cluster jet is surrounded by a sheath of He atoms that did not condense into clusters. By moving the cryostat so that the laser focus locates spatially either on the sheath of atoms (a) or on the cluster jet (b), one can easily identify the cluster contribution to the recorded XUV fluorescence signal. We like to add that both, the nanoplasma as well as the atomic target emit fluorescence in the visible and infrared spectral range. The long-wavelength emission in the direction of the laser beam propagation is blocked by the Al-filter in front of the XUV spectrometer in the present study in order to reduce the signal background. However, we like to note that laser-induced breakdown spectroscopy of helium gas was performed in the past to evaluate its spectral emission characteristics as well as the parameters of the formed plasma, namely electron temperature and electron density (see e.g.\,\cite{HANAFI200011}).

The radiative decay of irradiated atoms in the XUV spectral range leads to dominant fluorescence lines $3p \to 1s$ at 25.6\,nm, $4p \to 1s$ at 24.3\,nm, $5p \to 1s$ at 23.7\,nm, as well as $6p \to 1s$ at 23.4\,nm of He$^{+}$ ions\,\cite{doi:10.1063/1.3077727}. The vertical length of the fluorescence line in Fig.\,2a is given by the 1\,mm spatial resolution of the pinhole camera (limited by the aperture diameter) and folded with the atomic beam velocity and the corresponding fluorescence lifetime. The beam waist at the laser focus perpendicular to the dispersion direction of the XUV grating is much smaller (6\,$\mu$m). It limits the spectral resolution of the XUV spectrometer and is experimentally determined by the so-called knife-edge method.

In the case of fluorescence emitted from large He clusters the same transitions show up but with an overall higher signal strength and a slightly broadened line spectrum. In addition, also a new dominant contribution appears with a wide wavelength distribution that covers the spectral region from the $6p \to 1s$ fluorescence all the way to photon energies corresponding to the ionization potential of He$^+$ ions (22.78\,nm and 54.4\,eV, respectively) as depicted in Fig.\,\ref{fig:Z-Scan-Distribution}b. The additional broadband emission has a different vertical profile compared to the fluorescence lines emitted from a beam of atomic He (Fig.\,2a). On the first view, one can see that it originates from fluorescing ions emitting further downstream from the nozzle than the initial 3\,mm that separate the focus of the drive laser and the nozzle. The center of mass of the broadband emission, i.e. its source point is shifted to a larger distance of 3.3\,mm from the nozzle.

The origin of the new broadband emission is radiative transition of $n>6$ Rydberg states to the $1s$ state following the population via three-body recombination (TBR). Initially, the clusters are field-ionized by the intense laser pulse. During the outer ionization a positive cluster ion core is created, that prevents further outer ionization, while inner ionization continues. This leads to the trapping of quasifree electrons in illuminated He clusters. The created quasifree electrons are driven by the laser field releasing an avalanche of secondary electrons mainly by electron-impact ionization of the surrounding neutral atoms and single charged ions\,\cite{Medina_2023}. Note, the presence of ions in the vicinity of He atoms facilitates this ionization cascade, because it lowers their ionization threshold by 10-17\,eV\,\cite{Heidenreich2017}. Furthermore, stable He complexes around the ions are formed due to higher binding energy of an ion to the surrounding neutral He atoms compared to the interactions between neutral He atoms\,\cite{Atkins1959,PhysRevLett.29.214,Scheier2020}. These often called 'Atkins snowballs' will be highly mobile in the superfluid and quickly redistribute to minimize the total repulsion energy\,\cite{PhysRevLett.123.165301,PhysRevResearch.4.L022063}. 

In contrast to the positively charged ions building He$_{n}^{+}$ and He$_{n}^{2+}$ snowballs, whose density can exceed that of solid He, the laser-heated quasifree electrons are heliophobic and form extended void bubbles inside the droplets\,\cite{PhysRevLett.75.4079,Toennies1997,PhysRevLett.81.3892}. During the nanoplasma expansion driven by Coulomb and hydrodynamic forces the quasifree electrons cool down sufficiently for TBR to set in. According to the bottleneck model this occurs at an electron temperature where recombination happens predominantly into a small number of $n'$-quantum number states above $n=6$ with an ionization energy comparable to the kinetic energy of thermal quasifree electrons. These high-lying Rydberg states then radiatively decay into the $1s$-state. The photons from this decay show up on the camera spectrally above the $6p$-fluorescence line, which is only weakly populated upon interaction of intense IR pulses with the beam of He atoms. 

Also long-lived neutral He$^\star$ excitations in droplets\,\cite{Buchenau1991,PhysRevLett.87.153403,Haeften2011,Mauracher2018}, which are populated by the recombination of electrons with singly-charged He$^+$ cations, have been observed in the present experiments. Note, highly excited Rydberg states\,\cite{Haeften2005,PhysRevLett.106.083401}, autoionization\,\cite{PhysRevLett.91.043401,Asmussen_2021} and interatomic Coulombic decay processes\,\cite{Ovcharenko2020,PhysRevX.11.021011}, as well as electronic relaxation leading to the evaporation of Rydberg atoms and excimers (excited dimers) from the droplet surface\,\cite{Asmussen_2021,Kornilov2011,Mudrich2020,Haeften2023} have been studied previously. However, due to the limited performance of the installed XUV grating in the limit of long wavelengths and the presence of spectrally overlapping signals originating from the 2$^{nd}$ diffraction order, we did not evaluate the corresponding fluorescence spectra in much detail. Therefore, also no information is obtained here on the formation of metastable bubble states in He droplets \,\cite{PhysRevLett.71.700,Henne1998,PhysRevLett.81.3892,PhysRevLett.88.233401}, which nevertheless appears to be rather unlikely in the presence of ions in the laser-generated nanoplasma due to the strong Coulomb attraction. We argue that the high density of quasifree electrons inside fully ionized clusters combined with the enhanced scattering cross-section of the embedded He$_{n}^{2+}$ snowballs promotes three-body recombination into high-lying Rydberg states of the cations. In the case of illuminated He gas, no such quasifree electrons and snowballs exist. Here, the lack of local electron density maxima in the vicinity of laser-generated free ions results in a much reduced recombination rate into Rydberg states, which explains the absence of the corresponding spectrally broad emission band in Fig\,\ref{fig:Z-Scan-Distribution}a.
\begin{figure}[ht]
\includegraphics[width = 8.7cm]{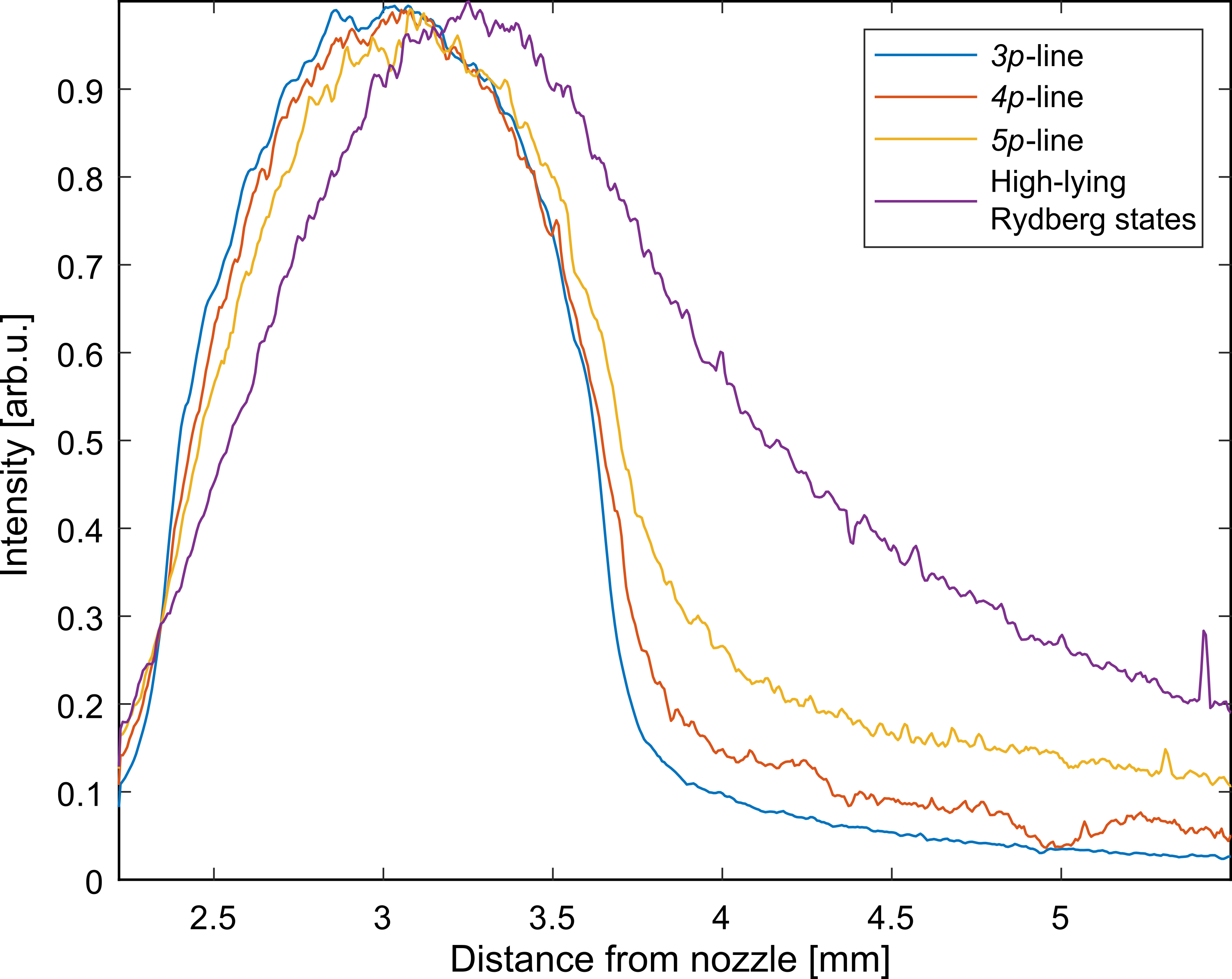}
\caption{\label{fig:long_tails_p-lines} Normalized vertical readout of the cluster spectrum (\mbox{$\overline{N}=1.2\times 10^6$}). The profiles were obtained by integrating the intensity of camera pixels, which record fluorescence light within the wavelength range of 25.84\,nm\,$<\lambda<$\,25.43\,nm (3p), 24.46\,nm\,$<\lambda<$\,24.16\,nm (4p), 23.88\,nm\,$<\lambda<$\,23.63\,nm (5p) and 23.59\,nm\,$<\lambda<$\,22.77\,nm for the broadband emission from high-lying Rydberg states}.
\end{figure}

The vertical displacement of the emission source point from high-lying Rydberg states and the tail of low-energy fluorescence lines towards larger nozzle distances share a common origin. It is due to cluster disintegration driven by Coulomb and hydrodynamic forces. Their rapidly increasing net charge in the course of massive energy deposition and outer ionization accompanied by the continuously increasing temperature of quasifree electrons drives the cluster expansion. This results in individual ions gaining up to several keV of kinetic energy. In between the population of low-energy $p$-states by frustrated tunneling ionization\,\cite{PhysRevLett.101.233001,Landsman_2013,PhysRevLett.110.203002,PhysRevA.94.013417,PhysRevLett.118.013003,Nature.12.620} or direct multi-photon excitation\,\cite{PhysRevLett.117.203001,Nature.11.252,PhysRevA.95.041401} during the laser-plasma interaction and its radiative transition to the $1s$-state an ion with keV kinetic energy moves a distance away from the IR laser focus. Accordingly, the length of the tail should scale with the lifetime of the electronically excited state, i.e. with the third power of the $n$ quantum number of the populated $p$-state. Note that corresponding tails in fluorescence lines emitted from a beam of atomic He (Fig. 2a) are missing, because compared to the cluster ion emitters, which move at high speed, an illuminated beam of atomic He emits fluorescence essentially in rest.

Qualitatively, we find exactly this behavior in our experiments as shown in Fig\,\ref{fig:long_tails_p-lines}, where the normalized vertical profiles of all detected fluorescence bands from illuminated clusters are plotted. The higher-lying $p$-state emission spectra do indeed show longer tails than the lower-lying ones. According to the literature, their lifetime vary from 374\,ps for the $3p$-state to 3.2\,ns for the $6p$-state\,\cite{doi:10.1063/1.3077727}. Thus, the electronically excited ions with keV kinetic energy from the disintegrating cluster have more time to move away from the nozzle before radiative decay takes place. We note in passing that the long tails of the lower-lying fluorescence lines, as well as the general shape of the high-lying broadband Rydberg emission are asymmetric with regards to the vertical axis, although the Coulomb explosion and hydrodynamic expansion that causes these dynamical processes is isotropic in nature. A possible and straight forward explanation is simply that closer towards the nozzle the ion and atom density markedly increases, which significantly enhances self-absorption. The effect might block the light emitted by ions that are pushed towards the nozzle by the Coulomb explosion resulting in the asymmetry we detect in Fig.\,\ref{fig:Z-Scan-Distribution}.

\begin{figure}[ht]
\includegraphics[width = 8.7cm]{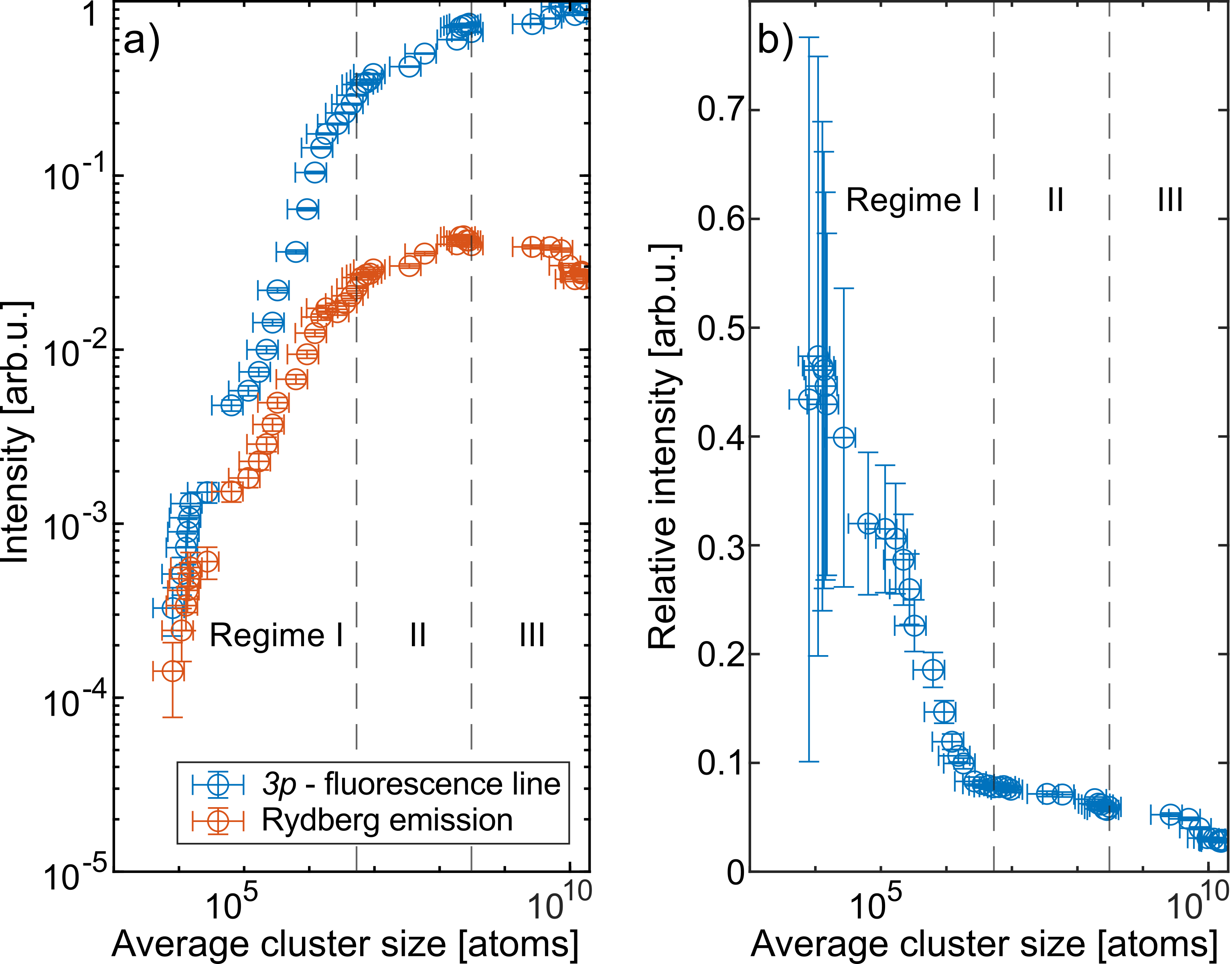}
\caption{\label{Absolute_yield_combined_cropped} a) Total yield of the $3p$-fluorescence line and the Rydberg emission recorded as a function of average cluster size $\overline{N}$. b) Relative intensity of the Rydberg emission compared to $3p$-fluorescence depending on $\overline{N}$. Dashed vertical lines indicate different regimes of cluster formation: supersonic gas expansion (I), cluster formation close to the critical point (II), and the formation of the largest droplets by fragmentation out of the liquid phase (III).}
\end{figure}

Next, we investigate the population of high-lying Rydberg states depending on the number of atoms per cluster in the range of $10^{4}<\overline{N}<10^{10}$. With the cluster jet located at the laser beam waist, the temperature of the cryostat was varied to change the particle size across the different regimes of cluster formation: supersonic gas expansion, cluster formation close to the critical point, and the formation of the largest He droplets by liquid He fragmentation \cite{Toennies2004}. As shown in Fig.\,\ref{Absolute_yield_combined_cropped}a, the total fluorescence yield of the Rydberg emission increases by more than four orders of magnitude with cluster size across the measured size range and the same holds true for the $3p$-fluorescence line. The dramatic signal increase is caused by the increased mass flow through the nozzle at low temperatures. Additionally, the opening angle of the He jet after leaving the nozzle decreases with lower temperatures and therefore with larger average cluster sizes. Both effects lead to a higher He density in the focus volume resulting in a stronger fluorescence signal. For the largest average droplet size the total fluorescence yield does not increase further. The reason is that light propagation effects become important within clusters measuring several hundred nanometers to a few micrometers in diameter. These droplets are too large to be fully ionized by the laser pulse and the emission is partially self-absorbed in the outer shells. Furthermore, the emission from inner parts of the droplets is quenched by collision-induced non-radiative decay processes mediated by the quasifree electrons. These arguments are based on theoretical works modelling the interaction between laser light and nanoparticles on microscopic length scales\,\cite{https://doi.org/10.1002/andp.201490001}. The studies take into account: (i) classical trajectories of all electrons and ions fully resolving microscopic processes such as collisions and (ii) capture wave propagation phenomena, which take place on the order of the laser wavelength. The analysis of these processes has become possible through the recent development of the microscopic particle-in-cell (MicPIC) approach\,\cite{NewJPhys.14.065011,PhysRevLett.108.175007}. The simulations show reduced MicPIC absorption for large clusters compared to electrostatic molecular dynamics simulations, which reflects radiation damping as a clear sign of propagation effects in the nanoplasma.

Interestingly, the relative signal strength of the Rydberg emission and the $3p$-fluorescence depends on the cluster size $N$ as displayed in Fig.\,\ref{Absolute_yield_combined_cropped}b. It shows a maximum for clusters comprising on average $10^{4}-10^{5}$ atoms, i.e. close to the minimum size needed ($N=10^{5}$) to support multiple charges in He clusters\,\cite{PhysRevLett.123.165301}, a plateau in the size range $10^{6}-10^{8}$ atoms and a decay for the largest droplets generated by fragmentation out of the liquid phase. Obviously, the relative population of high-lying Rydberg states exhibit a resonance behavior, which is similar to experimental observations and theoretical findings in previous nanoplasma studies on rare-gas clusters (see\,\cite{Saalmann2006} and references therein). It results from the dynamic interplay of cluster ionization, quasifree electron heating and nanoplasma expansion. The individual time-dependent rates depend on cluster size and laser parameters. During the resonance phase the cluster absorbs laser radiation extremely efficiently and rapidly heats up, which in turn accelerates its expansion rate. After the laser pulse has passed the cluster and any heating dynamics have stopped, the expansion of the cluster still continues. This leads to a steady reduction in the temperature of the electrons due to adiabatic cooling. Once the electron temperature has dropped far enough that electron-ion recombination sets in according to the bottleneck model, strong fluorescence from high-lying Rydberg states is observed. 

From the data we also learn that the population of high-lying Rydberg states in small He clusters comprising significantly less than $10^{4}$ atoms appears to be a rather rare event, since the quasifree electrons and ions rapidly expand and leave the clusters within less than 1 ps. The capability of larger He droplets to trap ions and electrons in stable complexes and quasifree states, respectively, thereby accumulating multiple charge carriers over longer periods of time, promotes three-body recombination in the laser-driven nanoplasma. Our findings complement recent studies on the XUV activation of He nanodroplets, where long-lasting changes of the strong-field optical properties of nanoparticles have been observed and attributed to electrons remaining loosely bound to photoions forming stable snowball structures in the droplets.\,\cite{Medina_2023}.    

\begin{figure}[ht]
\includegraphics[width = 8.7cm]{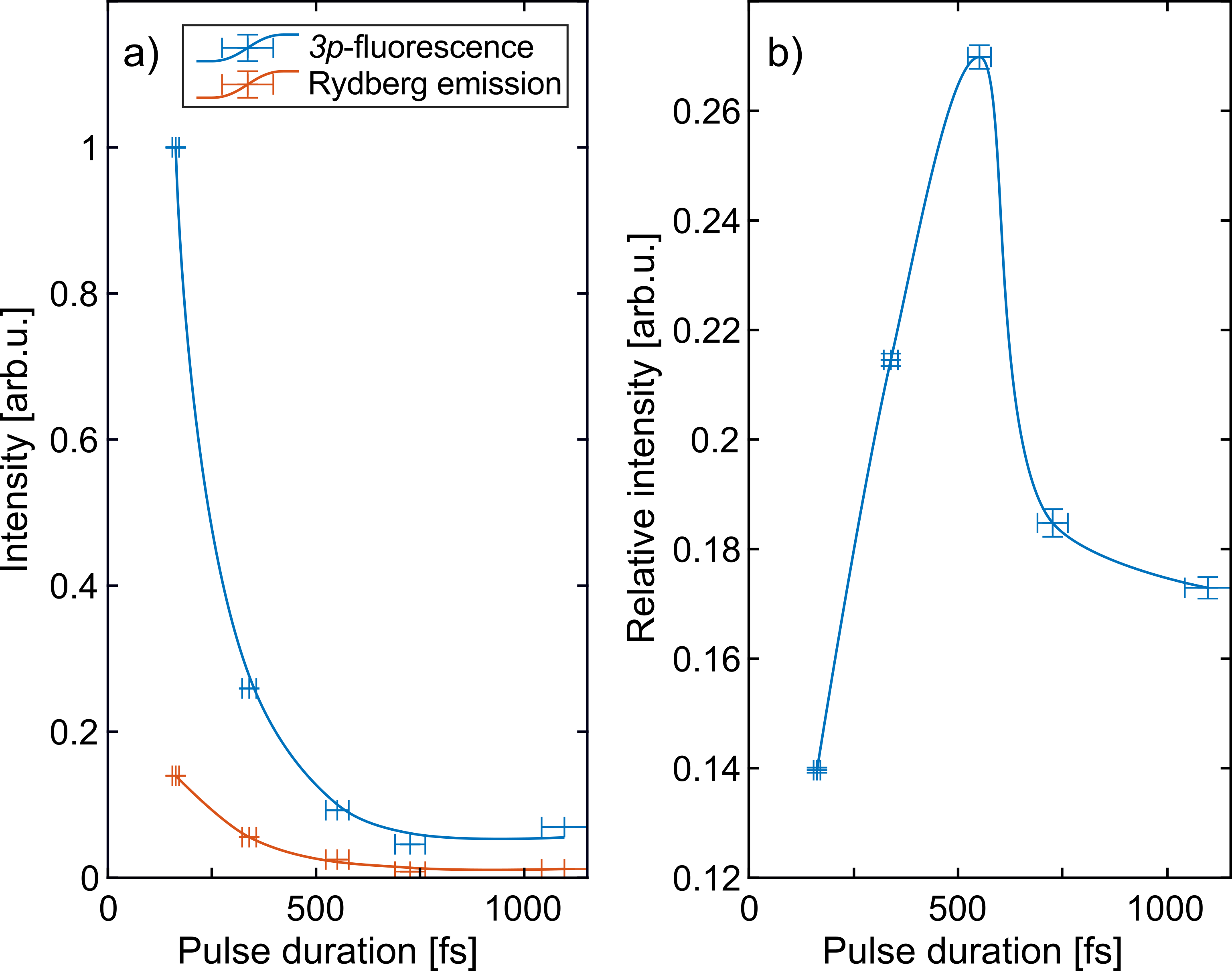}
\caption{\label{duration_925K_combined_cropped} (a) Total yield of the 3p-fluorescence line and the Rydberg emission as a function of laser pulse length for clusters consisting \mbox{$\overline{N}=1.2\times 10^6$} atoms and a pulse energy of 1\,mJ. (b) Ratio of the intensity of the Rydberg emission to the 3p-fluorescence line. Solid lines are added as guide to the eye.}
\end{figure}

As the heating of the nanoplasma depends sensitively on the laser pulse duration also the relative population of high-lying Rydberg states for a given cluster size should show some resonant character is this respect. Indeed, this has been observed in the present study. The recorded fluorescence yields of the $3p$-line and the Rydberg emission are plotted in Fig.\,\ref{duration_925K_combined_cropped}a as a function of the laser pulse duration. The experiments were carried out at a cluster size of $\overline{N}=1.2\times 10^6$ atoms varying the pulse duration between 180\,fs and 1.2\,ps. Depending on the setting of the grating compressor of the near-IR laser system the pulses contain different amounts of dispersion causing them to get stretched in time, which is characterized by recording intensity autocorrelation traces. 

One can clearly see that both the strength of the $3p$-fluorescence line and the Rydberg emission increases, when the clusters interact with pulses of the same energy but shorter duration. This is to be expected since shorter pulses with the same energy result in higher laser peak intensity and thus more efficient energy absorption. Note, the plasma heating rate increases with the ponderomotive potential and therefore also with the intensity of the drive laser. However, when plotting the relative strength of these emissions from electronically excited states populated in the transient nanoplasma as a function of laser pulse duration as done in Fig.\,\ref{duration_925K_combined_cropped}b the resonance is revealed. The emission from high-lying Rydberg states mediated by three-body recombination is strongest relative to the $3p$-fluorescence in the nanoplasma at a pulse duration of\,500 fs. Since in the present measurement the clusters are larger than those generating the strongest Rydberg $(>6p)$ emission with a pulse duration of 180\,fs (see Fig.\,\ref{Absolute_yield_combined_cropped}b), one expects slightly longer pulses to reach the resonant condition for efficient laser energy absorption. This is because, larger droplets turned into a nanoplasma expand more slowly as the inner part remains quasi-neutral and electrons and ion remain confined in the plasma for longer periods of time\,\cite{Krishnan_2012}. We note in passing that we recorded the total fluorescence yield of the $3p$-fluorescence line and the broadband Rydberg emission also as a function of laser pulse energy at a fixed pulse duration of 180\,fs for clusters comprising $\overline{N}=1.2\times 10^6$ atoms. Under these conditions the population of the electronically excited states scales linearly with the deposited energy into the nanoplasma (not shown here). 

Finally, we would like to briefly discuss the topic of coherence in the XUV nanoplasma emission induced by intense laser fields, which is of course a very interesting phenomena not only from an application point of view but also for basic research in attosecond physics. High-harmonic generation (HHG) in small nanoparticles of the heavier rare gases has been demonstrated in the past\,\cite{PhysRevLett.110.083902, JPhysB.30.L709,ALADI201668,10.1063/1.1888053,Bodi:19}. However, in the present study on large helium droplets exposed to strong laser fields we did not observe HHG, neither above the ionization threshold nor below, i.e. spectrally overlapping with the atomic Rydberg resonance energies. The likely reason is the use of rather long laser pulses with a duration of 180\,fs, which corresponds to 52 optical cycles at a wavelength of 1030\,nm. With such a large number of cycles it is likely that the cluster has been transformed into a helium nanoplasma before the peak intensity is reached, where efficient generation of XUV light typically occurs. It seems that once the transient nanoplasma is created the phase relation between the drive laser field and the generated XUV light in the dense heterogeneous medium is lost, which prevents the so-called phase matching condition in HHG to be fulfilled\,\cite{Science.280.1412}.

In the case of the strong-field interaction with a beam of atomic He (Fig.\,2a), given the wavelength $\lambda = 1030$\,nm and the focal spot size of 6\,$\mu$m, we obtain a minimum phase matching pressure 4.5\,mbar. Considering that the gas nozzle has a diameter of 5\,$\mu$m and we apply a pressure of 20\,bar and assuming an opening angle of 45\,$^{\circ}$ for the gas jet, we obtain the 4.5\,mbar at a distance from the nozzle of 166\,$\mu$m. This is much closer to the nozzle than the 3\,mm used in the experiment. Therefore, we could also not achieve phase matching and observe coherent HHG emission in the beam of atomic He (Fig.\,2a). Nevertheless, we would like to emphasize that the population of Rydberg states in atoms by intense femtosecond laser pulses is a coherent process and all atoms in the gas medium decay in phase to the ground state by spontaneous emission, resulting in coherent emission typically referred to as XUV free induction decay (XFID)\,\cite{PhysRevLett.117.203001}. 

In atoms it has been shown that the coherent emission of Rydberg states in the extreme ultraviolet spectral range can also be populated by recapture of the tunneling electrons during the interaction with the laser field in addition to resonant multi-photon excitation\,\cite{Nature.12.620}. This is very similar to HHG, because recombination to the excited state occurs after excursion along electron trajectories generating XFID pulses of nanosecond duration. It would be extremely interesting to study the temporal properties of the XUV nanoplasma emission by using for instance the so-called attosecond lighthouse technique\,\cite{PhysRevLett.108.113904,Nature.7.651}. The attosecond lighthouse maps the emission time onto the propagation direction and therefore provides important information on the temporal confinement of the emission from Rydberg states. However, the application of this advanced technology developed by the attosecond science community using atomic targets goes beyond the scope of the present nanoplasma study.

In summary, utilising fluorescence spectroscopy to investigate energy absorption and dissipation in helium clusters and large droplets irradiated with intense femtosecond laser pulses at central wavelength of 1030\,nm revealed a number of interesting phenomena. First of all, it is clearly observed, how the dense cluster environment opens up an additional relaxation pathway in the laser-generated nanoplasma via three-body recombination (TBR) in comparison to the laser interaction with a beam of He atoms. The nanoplasma-specific relaxation pathway results in the population of high-lying Rydberg states of He ions upon plasma expansion and adiabatic electron cooling, respectively. The fluorescence decay into the ionic ground state results in a characteristic broadband emission of XUV photons with energies close to the ionization potential of He$^+$. This finding is in agreement with previous studies using electron and ion spectroscopy to unravel the role of Rydberg states in the energetics and dynamics of laser-generated nanoplasma. The XUV spectrometer setup in the present work, which also acts as a pinhole camera, allowed to make qualitative statements about the relative time frame of the TBR relaxation pathway by a time to space mapping of the fluorescence source point in the expanding nanoplasma. This is because the cluster has to expand, i.e. the electrons need to thermalize before high-lying Rydberg-states are efficiently populated by low-energy quasifree plasma electrons according to the well-known bottleneck model. By varying the size of the clusters, we were able to study the size-dependence of the resonant plasma heating, since the TBR relaxation pathway is much more sensitive to the temperature of quasifree electrons than is fluorescence from direct laser excitation of low-lying electronic states also populated in the plasma. Last but not least, we were able to corroborate the electron-temperature sensitivity of TBR by influencing the resonant heating condition applying laser pulses of different duration.

\section*{Acknowledgement}
This research was supported by the German Federal Ministry for Economic Affairs and Climate Action through the collaborative ZIM research project KK5322301DF1.

\bibliography{Helium_Cluster_Bib.bib}

\end{document}